# Strong Oxidation Resistance of Atomically Thin Boron Nitride Nanosheets

*Lu Hua Li,[†]\* Jiri Cervenka,[‡] Kenji Watanabe,[§] Takashi Taniguchi[§] and Ying Chen[†]\**

[†]Institute for Frontier Materials, Deakin University, Geelong Waurn Ponds Campus, Victoria 3216, Australia

[‡]School of Physics, The University of Melbourne, Victoria 3010, Australia

[§]National Institute for Materials Science, Namiki 1-1, Tsukuba, Ibaraki 305-0044, Japan

ABSTRACT

Investigation on oxidation resistance of two-dimensional (2D) materials is critical for many of their applications, because 2D materials could have higher oxidation kinetics than their bulk counterparts due to predominant surface atoms and structural distortions. In this study, the oxidation behavior of high-quality boron nitride (BN) nanosheets of 1-4 layer thick has been examined by heating in air. Atomic force microscopy and Raman spectroscopy analyses reveal that monolayer BN nanosheets can sustain up to 850 °C and the starting temperature of oxygen doping/oxidation of BN nanosheets only slightly increases with the increase of nanosheet layer and depends on heating conditions. Elongated etch lines are found on the oxidized monolayer BN nanosheets, suggesting that the BN nanosheets are first cut along the chemisorbed oxygen chains and then the oxidative etching grows perpendicularly to these cut lines. The stronger oxidation resistance of BN nanosheets suggests that they are more preferable for high-temperature applications than graphene.

Atomically thin boron nitride (BN) nanosheets, a structure analogue of graphene, have superior mechanical and thermal conducting properties, and thus been used as reinforcing fillers in composites to improve their mechanical and thermal performances.[1,2] In contrast to zero bandgap graphene, BN nanosheets are a wide bandgap semiconductor suitable for optoelectronics[3,4] or as dielectric substrate for high-performance graphene electronics.[5-7] Graphene sandwiched by monolayer BN is predicted to have a tunable bandgap without sacrificing its mobility.[8] However, many intrinsic properties of few-layer BN nanosheets have not been experimentally investigated. For example, it has been found that BN nanosheets have a strong tendency to adsorb organic contaminations from both atmosphere and lithographic process,[9,10] which may affect the function of the graphene electronics using BN nanosheets as dielectric substrates as well as the bandgap opening of graphene sandwiched by monolayer BN. Heating in oxygen containing gas at 500 °C has to be used to remove the contamination.[9] However, it is unknown whether this heating treatment can oxidize the BN nanosheets and introduce defects/pinholes, which might lead to current leakage in the case of BN-substrated graphene electronics. Although bulk hBN crystals, multiwalled BN nanotubes and BN nanosheets show strong thermal stability,[11-14] monolayer thin BN may have much lower oxidation resistance, similar to the case of graphene. The stability of graphene in air shows a clear thickness dependence: monolayer graphene is reactive to oxygen at 250 °C, strongly doped at 300 °C and etched at 450 °C; in contrast, bulk graphite is not oxidized till 800 °C.[15] Therefore, it is critical to investigate the oxidation behavior of atomically thin BN nanosheets.

RESULTS AND DISCUSSION

In this work, the stability of high-quality free-standing monolayer and few-layer BN nanosheets at different temperatures in air is evaluated. The BN nanosheets were exfoliated on Si wafer covered with a 90 nm thick oxide layer ($SiO_2$/Si) from single crystal hBN[16] using the Scotch tape method.[17] The few-layer BN nanosheets were identified using an optical microscope and then the thickness was measured using an atomic force microscope (AFM) . Our AFM results show that monolayer BN on $SiO_2$/Si normally has a height of 0.4-0.5 nm, and bilayer and trilayer of 0.7-0.9 and 1.1-1.3 nm, respectively. These results are consistent with the previously reported AFM measurements on few-layer BN.[17] Figure 1a shows an AFM image of a monolayer BN, whose corresponding height trace and optical photo are present in Figure 1b. In the AFM image, the brighter the color is, the larger the thickness is the nanosheet, and vice versa. The AFM and optical images of bilayer and trilayer BN nanosheets can be found in Figure S1 in Supporting Information.

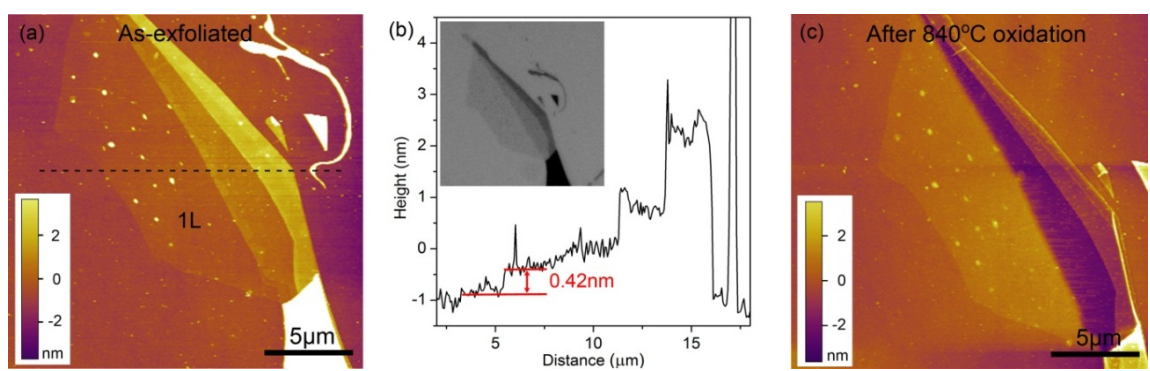

Figure 1. (a) AFM image of a monolayer BN before heating, along with some tape residues from the exfoliation process (top-right); (b) the height trace of the dashed line in (a) and the corresponding optical microscope photo inserted; (c) AFM image of the same monolayer BN after 840 °C heating in air for 2 h.

The oxidation tests were conducted by heating the atomically thin BN and bulk crystals in a tube furnace at different temperatures (400, 500, 600, 700, 800, 840, 850, 860 and 870 °C) with a holding time of 2 h in open air. No noticeable morphology change or etching has been observed on any BN nanosheets after 2 h heating up to 840 °C in AFM, indicating that BN nanosheets are quite stable to oxidation. Figure 1a and 1c display AFM images of the same monolayer BN before and after the 840 °C heat treatment. Monolayer BN starts to show oxidative etch pits at 850 °C (Figure 2a) and burns out completely at 860 °C after 2 h heating. Bilayer BN starts to be etched at a higher temperature of 860 °C (Figure 2d) and burns out at 870 °C; while trilayer BN only partially burns at 870 °C. The oxidation behaviour of the tested BN nanosheets has not been found lateral size dependent, and BN nanosheets do not seem to have reaction with the $SiO_2$ substrate up to 870 °C consistent with the HSC thermodynamic calculation. Most bulk hBN crystals are not affected by heating up to 870 °C under the same condition, though etch pits and lines are occasionally found on their surface (see Supporting Information, Figure S3).

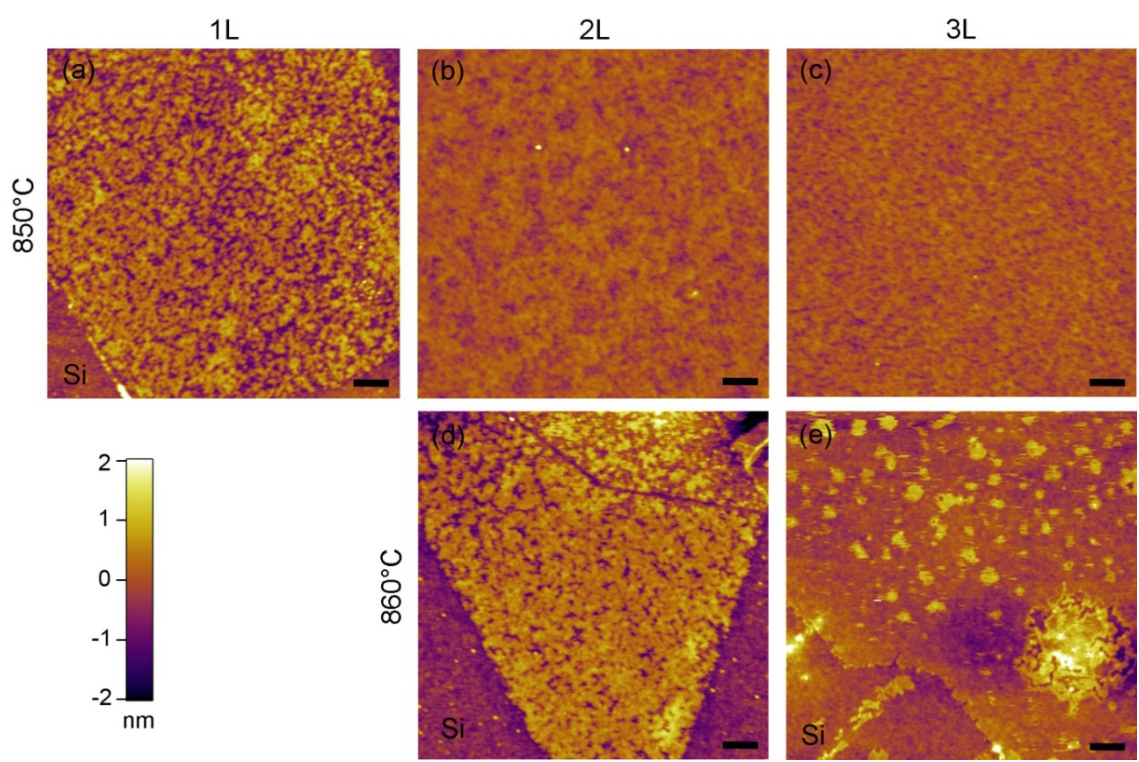

Figure 2. AFM images of monolayer (1L), bilayer (2L) and trilayer (3L) BN nanosheets after 850 and 860 °C heating in air for 2 h. Monolayer BN burns out after 860 °C heating. All scale bars are 200 nm.

Direct quantitative measurement on the oxygen contents in these oxidized monolayer BN is challenging, but Raman spectroscopy can provide detailed information on the oxidation progress of atomically thin BN nanosheets. Raman spectra of BN nanosheets are different from those of graphene, showing only a Raman G band corresponding to $E_{2g}$ vibration mode[17] but no D band due to the lack of Kohn anomaly (Figure 3a). Figure 3b summarizes the G band frequency ($\omega_G$) of the atomically thin BN of different layers without any oxidation treatment. The G band frequency of 1-4 layer BN is upshifted compared to that of bulk sample. The G band frequencies are 1366.2±0.2 cm$^{-1}$ for bulk BN samples (average deviation from 6 samples, $N$=6), 1370.5±0.8 cm$^{-1}$ for monolayer ($N$=15), 1370.0±0.6 cm$^{-1}$ for bilayer ($N$=10), 1367.8±0.4 cm$^{-1}$ for trilayer ($N$=2) and 1367.2±0.4 cm$^{-1}$ for 4-layer BN ($N$=2). The G band upshift for the BN nanosheets is attributed to their higher level of in-plane strain and lower interlayer interaction,[18] both leading to phonon softening. The 150 °C vacuum heating treatment has not caused any noticeable G band shift. In addition to the different G band frequency, the full widths at half maximum (FWHMs) of the G band ($\Gamma_G$) of monolayer and bilayer BN are dramatically larger than that of the bulk sample (Figure 3c). This is due to larger strain arising from the interaction with the substrate and stronger surface scattering in atomically thin BN,[17,19] both of which can influence the vibrational excitation lifetime and thus increase the Raman band width. The observed Raman frequency shifts are not in line with a previous report on BN nanosheets exfoliated from single crystal hBN, which shows G band upshift for monolayer and downshift for 2-6 layer BN.[17] Reason for this discrepancy is currently unknown and needs further investigation.

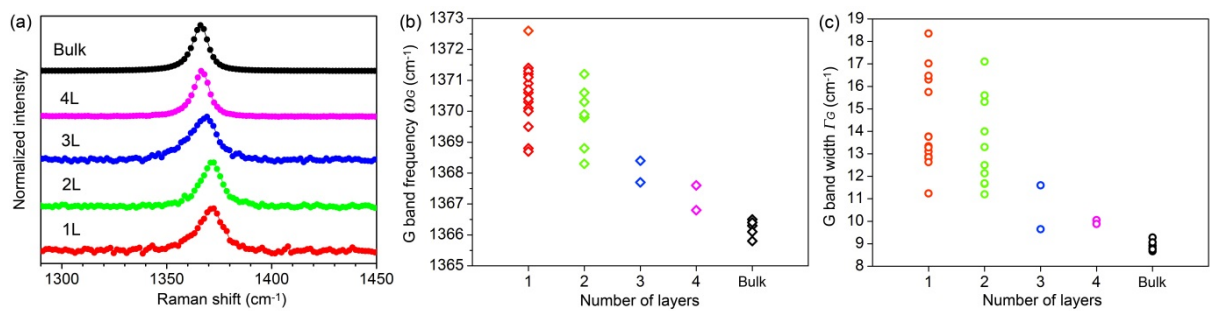

Figure 3. (a) Normalized Raman spectra of 1-4 layer BN nanosheets and bulk hBN on SiO$_2$/Si substrate before oxidation; (b) corresponding Raman G band frequencies and (c) widths determined from Voigt curve fitting. The Raman frequency was calibrated using the 520.5 cm$^{-1}$ peak of the Si substrate.

The Raman frequency shift and width (FWHM) change of the BN nanosheets after 2 h heating in air at different temperatures are summarized in Figure 4. The G band intensity of monolayer BN starts to decrease at 800 °C and dramatically weakens at 850 °C (1L in Figure 4a). Though a complete theory for the intensity of the G band of BN is still lacking, the observed intensity reduction after 800 °C heating can be due to a decreased number of B–N bonds satisfying Raman selection rule, which might imply an increased degree of oxidation. The average G band frequency and width changes of monolayer BN as a function of heating temperature are plotted in Figure 4b and 4c (1L). It can be seen that the G band of monolayer BN broadens after 800 °C heating. As the broadening can be caused by oxygen doping, crystal domain size change, disordering or amorphization,[15,19] it indicates that oxygen doping and adsorption on monolayer BN starts at 800 °C. The G band broadening is consistent with the observed G band intensity decrease at the same temperature (1L in Figure 4a). However, the oxygen doping and adsorption below 840 °C does not cause any dramatic surface morphology change of the BN nanosheets in AFM, as shown in Figure 1c. Bilayer and trilayer BN show decreases of the G band intensity (2L and 3L in Figure 4a) and increases in the G band frequency and width (2L and 3L in Figure 4b and c) after 860 °C oxidation, revealing their slightly higher oxidation temperatures than the monolayer. The beginning temperature of oxidation for 4-layer BN is 870 °C (4L in Figure 4). It should be noted that the G band broadenings of the atomically thin BN oxidized at the 850-870 °C are dramatically lower than those observed in hBN materials with a high degree of disorder or amorphization (whose $\Gamma_G$ could be up to 400 cm$^{-1}$).[20] Because the G band width quantitatively correlates with the crystal domain size due to a wave-vector uncertainty, the oxygen doping concentration can be roughly estimated from the G band broadening if the oxygen doping sites are considered as domain boundaries in BN nanosheets. By using the formula $L_a = 1417/\Gamma_G - 8.7$,[19] where $L_a$ is domain size in Å, the calculated domain sizes for 1-4 layer BN nanosheets at 850-870 °C are around 20 nm and the estimated oxygen doping concentration is in the order of $10^{-5}$ at.%. In contrast, the bulk hBN has almost no G band shift and broadening after 870 °C heating (Bulk in Figure 4), indicating very little oxidization. So it can be concluded that atomically thin BN nanosheets have lower oxidation resistance than the bulk crystals, but overall are much more stable than graphene. Although the oxidation of monolayer, bilayer and trilayer BN are thickness dependent, the exact oxidation temperatures are quite close (~10 °C).

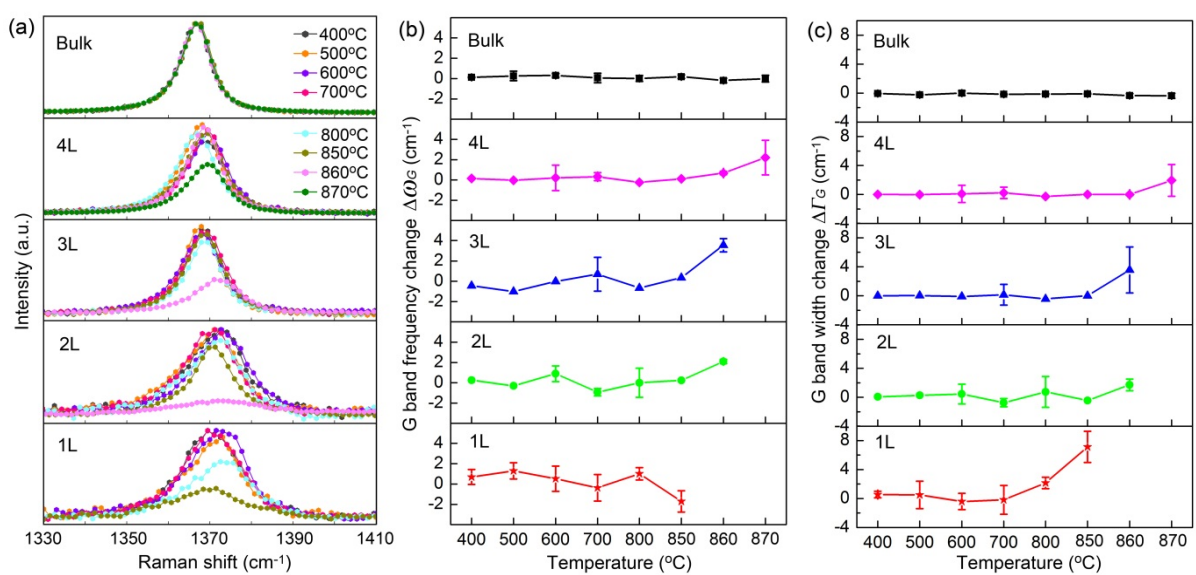

Figure 4. (a) Raman spectra of 1-4 layer thick BN nanosheets and bulk hBN after oxidation at different temperatures for 2 h; (b) the corresponding G band frequency and (c) width changes.

The oxidation resistance of atomically thin BN nanosheets was also tested by repeated heating of the same pieces of nanosheets in sequence at 400, 500, 600, 700 and 800 °C in air for 2 h at each temperature (*i.e.* heated at 400 °C for 2 h and then 500 °C for 2 h and so on). There are mainly two reasons for conducting these experiments: (1) repeated heating (including cooling) will mimic a more realistic heat treatment as in many applications; (2) it can be considered as extended heating treatment (*e.g.* the total oxidation time from 400 to 800 °C sequential heating is 10 h). Figure 5 shows the G band frequency and width of the sequentially heated atomically thin BN. The G band intensity of the monolayer BN dramatically reduces at 700 °C (1L in Figure 5a) and the monolayer BN burns out at 800 °C in the sequential heating. Both temperatures are lower than those of the samples heated once for 2 h (Figure 4). The 2-4 layer BN nanosheets show some oxidation after repeated heating up to 700 °C and are severely oxidized at 800 °C, as evidenced by the decreased G band intensities (2L, 3L and 4L in Figure 5a) and the increased G band widths (2L, 3L and 4L in Figure 5c). In addition, the tri- and 4-layer BN show relatively strong G band upshift after 700 °C sequential heating (3L and 4L in Figure 5b), suggesting higher levels of oxygen doping. The estimated $L_a$ for the bilayer and trilayer BN after repeated heating up to 800 °C is around 10 nm and the oxygen doping concentration is also in the order of $10^{-5}$ at.%. Similar to the case of graphene,[21] repeated and extended heating treatment can decrease the burn-out temperature and increase the doping concentrations of BN nanosheets. The reduced stability after repeated heating in air may relate to increased roughness (or surface curvature) and possible bonding distortions in BN nanosheets, both of which can give rise to lower activation energy for oxidation.[15] The bulk hBN, on the other hand, does not seem to be affected by the repeated heating.

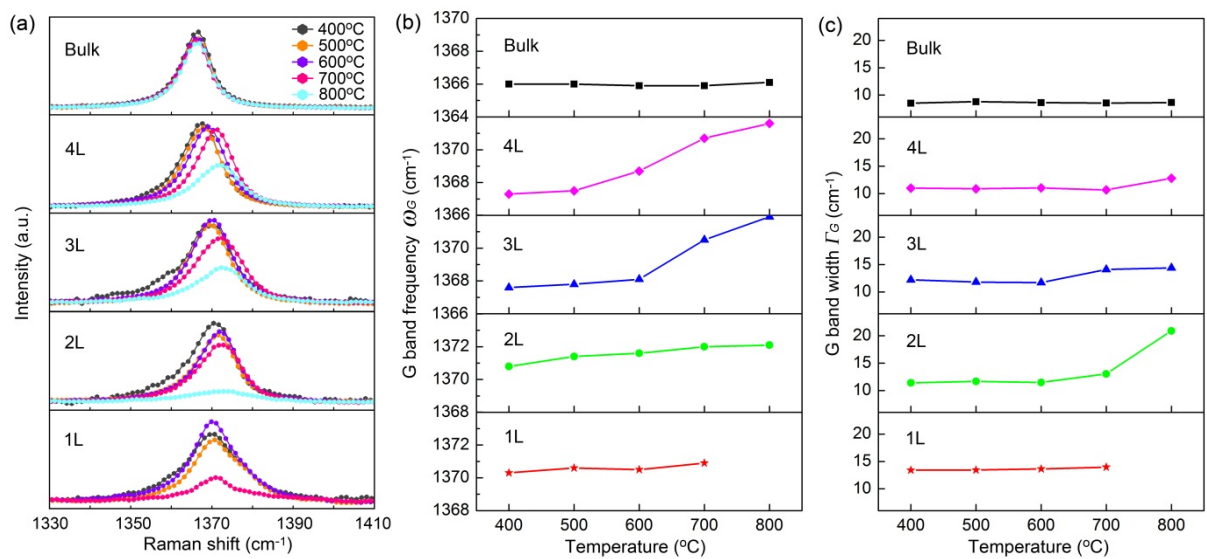

Figure 5. (a) Raman spectra of 1-4 layer BN nanosheets and bulk hBN after sequential heating at 400, 500, 600, 700 and 800 °C (2 h at each temperature); (b) the corresponding G band frequency and (c) width. The monolayer BN burns out after 800 °C heating in air.

The oxidative etching pattern on BN nanosheets is very different from that of graphene, indicating a different oxidation mechanism. Relatively uniform etch pits have been observed on oxidized graphene due to the radial oxidation from the pre-existing point defects.[15] Few of our atomically thin BN nanosheets showed such etch pits, probably because they were exfoliated from single crystal with few defects. After the oxidation treatment, elongated and randomly oriented etch pits were observed on the atomically thin BN nanosheets (several to more than 100 nm in width and 30 to 600 nm in length) (Figure 6). Theoretical calculations on the oxidation of monolayer BN predict that oxygen atoms are first chemisorbed on top of BN planes to form energy favorable oxygen chains and then the dissociation of the B–O–N bonds cuts the BN nanosheets straight along the chains.[22-25] Our experimental results support these predicted oxidation mechanism because the observed elongated etch lines on the oxidized BN nanosheets are mostly likely to be due to the further oxidation and broadening of the straight cut lines. In addition, oxygen doping can be formed in the oxidized BN nanosheets in the form of substitution of the nitrogen atoms by one, two or three oxygen atoms in the planar structure.[26-28] Atomically thin BN nanosheets have a strong oxidation resistance because energy barrier of molecular oxygen adsorption is higher than the desorption energy,[23] as shown by our Raman results that oxygen doping is not present on monolayer and other atomically thin BN till 700 °C heating in air.

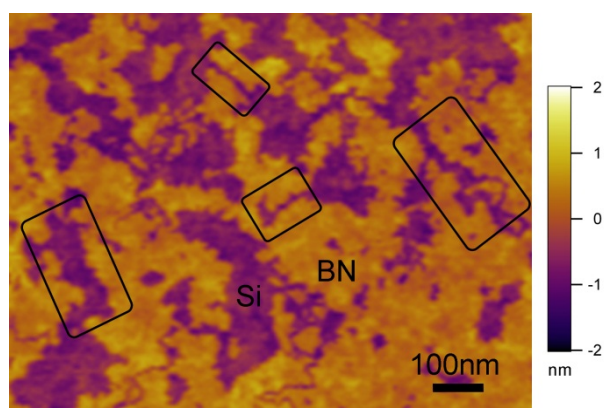

Figure 6. AFM image of an oxidized monolayer BN (yellow) on SiO$_2$/Si (purple), with elongated etch paths.

CONCLUSION

In summary, the oxidation resistance of high-quality and atomically thin BN nanosheets has been examined by heating in air at temperatures in the range of 400-870 °C under two conditions. Monolayer BN starts to oxidize at 700 °C and can sustain temperature up to 850 °C. Bilayer and trilayer BN nanosheets have slightly higher oxidation starting temperatures. The elongated oxidative etching observed in AFM suggests that the BN nanosheets are oxidized *via* the oxygen chain chemisorption and cutting mechanism. Compared to the 250 °C oxidation temperature of graphene, atomically thin BN nanosheets are more resistant to oxidation and therefore more suitable for high-temperature applications. Our results suggest that the heating of high-quality BN nanosheets in oxygen containing gas below 600 °C to remove adsorbed contaminations should not introduce defects or pinholes to the BN nanosheets.

EXPERIMENTAL SECTION

Single crystal hBN was used to exfoliate the BN nanosheets on 90 nm $SiO_2$/Si substrates following the Scotch tape method. The few-layer BN nanosheets were identified using an Olympus BX51 optical microscope equipped with a DP71 camera. A Cypher (Asylum Research) AFM with silicon cantilevers (spring constant 7.4 N/m, NanoWorld) operated in tapping mode was used to measure the thickness of the BN nanosheets after the samples heated in vacuum ($10^{-4}$ mbar) at 150 °C for 1 h to remove possible moisture adsorption. The temperature of the furnace used for the heating of the BN nanosheets had been calibrated by a thermal couple. Two confocal micro-Raman spectrometers (both inVia, Renishaw) equipped with 514.5 nm lasers (with maximum powers of 20 and 50 mW) were used to characterize the BN nanosheets before and after oxidation treatment.

ASSOCIATED CONTENT

**Supporting Information**

Additional AFM images of pristine BN nanosheets before heating treatment, AFM images of a monolayer BN after 800 °C heating in air for 2 h and AFM images of slightly oxidized bulk hBN. This material is available free of charge *via* the Internet at http://pubs.acs.org.

AUTHOR INFORMATION

**Corresponding Author**

*E-mail: luhua.li@deakin.edu.au; ian.chen@deakin.edu.au

**Notes**

The authors declare no competing financial interest.

ACKNOWLEDGMENT

L. H. Li thanks the kind help from Prof. Konstantin Novoselov from University of Manchester and Dr. Peter Blake from Graphene Industries in mechanical exfoliation procedure. The research project is supported in part by a Discovery grant from Australian Research Council.